\pdfoutput=1
\documentclass[12pt,a4paper]{article}
\usepackage{ifthen} % for conditional statements
\newboolean{pdflatex}
\setboolean{pdflatex}{true} % False for eps figures 

\newboolean{articletitles}
\setboolean{articletitles}{true} % False removes titles in references

\newboolean{uprightparticles}
\setboolean{uprightparticles}{false} %True for upright particle symbols

\def\paperauthors{LHCb collaboration} % Leave as is for PAPER, CONF and FIGURE
\def\paperasciititle{Search for the doubly heavy baryons Xi_bc and Omega_bc decaying to Lc-pi and Xic-pi} % Set ASCII title here !! MAKE sure it's only ASCII characters !! 
\def\papertitle{Search for the doubly heavy baryons \Omegabcn and \Xibcn decaying to $\Lcp \pim$ \\ and $\Xicp \pim$} % Latex formatted title
\def\paperkeywords{{High Energy Physics}, {LHCb}} % Comma separated list
\def\papercopyright{\the\year\ CERN for the benefit of the LHCb collaboration} % new since 9/Apr/2018
\def\paperlicence{CC BY 4.0 licence}
\def\paperlicenceurl{https://creativecommons.org/licenses/by/4.0/}

\usepackage[top=1in, bottom=1.25in, left=1in, right=1in]{geometry}

\usepackage{microtype}
\usepackage{lineno}  % for line numbering during review
\usepackage{xspace} % To avoid problems with missing or double spaces after
\usepackage{caption} %these three command get the figure and table captions automatically small

\usepackage{graphicx}  % to include figures (can also use other packages)
\usepackage{color}
\usepackage{colortbl}
\graphicspath{{./figs/}} % Make Latex search fig subdir for figures
\usepackage{amsmath} % Adds a large collection of math symbols
\usepackage{amssymb}
\usepackage{amsfonts}
\usepackage{upgreek} % Adds in support for greek letters in roman typeset

\newcommand*\patchAmsMathEnvironmentForLineno[1]{%
\expandafter\let\csname old#1\expandafter\endcsname\csname #1\endcsname
\expandafter\let\csname oldend#1\expandafter\endcsname\csname
end#1\endcsname
 \renewenvironment{#1}%
   {\linenomath\csname old#1\endcsname}%
   {\csname oldend#1\endcsname\endlinenomath}%
}
\newcommand*\patchBothAmsMathEnvironmentsForLineno[1]{%
  \patchAmsMathEnvironmentForLineno{#1}%
  \patchAmsMathEnvironmentForLineno{#1*}%
}
\AtBeginDocument{%
\patchBothAmsMathEnvironmentsForLineno{equation}%
\patchBothAmsMathEnvironmentsForLineno{align}%
\patchBothAmsMathEnvironmentsForLineno{flalign}%
\patchBothAmsMathEnvironmentsForLineno{alignat}%
\patchBothAmsMathEnvironmentsForLineno{gather}%
\patchBothAmsMathEnvironmentsForLineno{multline}%
\patchBothAmsMathEnvironmentsForLineno{eqnarray}%
}

\usepackage{hyperxmp}

\usepackage[pdftex,
            pdfauthor={\paperauthors},
            pdftitle={\paperasciititle},
            pdfkeywords={\paperkeywords},
            pdfcopyright={Copyright (C) \papercopyright},
            pdflicenseurl={\paperlicenceurl}]{hyperref}
\usepackage[colorinlistoftodos,textsize=scriptsize]{todonotes}

\usepackage[bottom,flushmargin,hang,multiple]{footmisc}

\usepackage[all]{hypcap} % Internal hyperlinks to floats.

\usepackage{xspace} 
\usepackage{upgreek}

\def\lhcb   {\mbox{LHCb}\xspace}

\def\MagUp {\mbox{\em Mag\kern -0.05em Up}\xspace}

\ifthenelse{\boolean{uprightparticles}}%
{

 \def\Ppi         {\ensuremath{\uppi}\xspace}

 \def\PDelta      {\ensuremath{\Delta}\xspace}                 
 \def\PXi         {\ensuremath{\Xi}\xspace}                 
 \def\PLambda     {\ensuremath{\Lambda}\xspace}                 
 \def\PSigma      {\ensuremath{\Sigma}\xspace}                 
 \def\POmega      {\ensuremath{\Omega}\xspace}                 
 \def\PUpsilon    {\ensuremath{\Upsilon}\xspace}

 \def\PB      {\ensuremath{\mathrm{B}}\xspace}                 
                  
 \def\PD      {\ensuremath{\mathrm{D}}\xspace}

 \def\PK      {\ensuremath{\mathrm{K}}\xspace}

 \def\Pb      {\ensuremath{\mathrm{b}}\xspace}                 
 \def\Pc      {\ensuremath{\mathrm{c}}\xspace}                 
 \def\Pd      {\ensuremath{\mathrm{d}}\xspace}

 \def\Pi      {\ensuremath{\mathrm{i}}\xspace}

 \def\Pp      {\ensuremath{\mathrm{p}}\xspace}

 \def\Ps      {\ensuremath{\mathrm{s}}\xspace}                 
                  
 \def\Pu      {\ensuremath{\mathrm{u}}\xspace}

 \def\thebaroffset{0.0em}
}
{

 \def\Ppi         {\ensuremath{\pi}\xspace}

 \mathchardef\PDelta="7101
 \mathchardef\PXi="7104
 \mathchardef\PLambda="7103
 \mathchardef\PSigma="7106
 \mathchardef\POmega="710A
 \mathchardef\PUpsilon="7107
                  
 \def\PB      {\ensuremath{B}\xspace}                 
                  
 \def\PD      {\ensuremath{D}\xspace}

 \def\PK      {\ensuremath{K}\xspace}

 \def\Pb      {\ensuremath{b}\xspace}                 
 \def\Pc      {\ensuremath{c}\xspace}                 
 \def\Pd      {\ensuremath{d}\xspace}

 \def\Pi      {\ensuremath{i}\xspace}

 \def\Pp      {\ensuremath{p}\xspace}

 \def\Ps      {\ensuremath{s}\xspace}                 
                  
 \def\Pu      {\ensuremath{u}\xspace}

 \def\thebaroffset{0.18em}
}
\newcommand{\offsetoverline}[2][\thebaroffset]{\kern #1\overline{\kern -#1 #2}}%

\makeatletter
\ifcase \@ptsize \relax% 10pt
  \newcommand{\miniscule}{\@setfontsize\miniscule{4}{5}}% \tiny: 5/6
\or% 11pt
  \newcommand{\miniscule}{\@setfontsize\miniscule{5}{6}}% \tiny: 6/7
\or% 12pt
  \newcommand{\miniscule}{\@setfontsize\miniscule{5}{6}}% \tiny: 6/7
\fi
\makeatother

\DeclareRobustCommand{\optbar}[1]{\shortstack{{\miniscule (\rule[.5ex]{1.25em}{.18mm})}
  \\ [-.7ex] $#1$}}

\def\uquark    {{\ensuremath{\Pu}}\xspace}

\def\dquark    {{\ensuremath{\Pd}}\xspace}

\def\squark    {{\ensuremath{\Ps}}\xspace}

\def\cquark    {{\ensuremath{\Pc}}\xspace}

\def\bquark    {{\ensuremath{\Pb}}\xspace}

\def\pion   {{\ensuremath{\Ppi}}\xspace}

\def\pip    {{\ensuremath{\pion^+}}\xspace}
\def\pim    {{\ensuremath{\pion^-}}\xspace}

\def\kaon    {{\ensuremath{\PK}}\xspace}
\def\KorKbar {\kern \thebaroffset\optbar{\kern -\thebaroffset \PK}{}\xspace}

\def\Kp      {{\ensuremath{\kaon^+}}\xspace}
\def\Km      {{\ensuremath{\kaon^-}}\xspace}

\def\D       {{\ensuremath{\PD}}\xspace}

\def\DorDbar {\kern \thebaroffset\optbar{\kern -\thebaroffset \PD}\xspace}

\def\Dp      {{\ensuremath{\D^+}}\xspace}
\def\Dm      {{\ensuremath{\D^-}}\xspace}

\def\DpDm    {\ensuremath{\Dp {\kern -0.16em \Dm}}\xspace}

\def\Dsp     {{\ensuremath{\D^+_\squark}}\xspace}

\def\B       {{\ensuremath{\PB}}\xspace}

\def\BorBbar {\kern \thebaroffset\optbar{\kern -\thebaroffset \PB}\xspace}

\def\Bd      {{\ensuremath{\B^0}}\xspace}

\def\BdorBdbar {\kern \thebaroffset\optbar{\kern -\thebaroffset \Bd}\xspace}

\def\Bs      {{\ensuremath{\B^0_\squark}}\xspace}

\def\BsorBsbar {\kern \thebaroffset\optbar{\kern -\thebaroffset \Bs}\xspace}

\def\Y#1S{\ensuremath{\PUpsilon{(#1S)}}\xspace}

\def\proton      {{\ensuremath{\Pp}}\xspace}

\def\Lz          {{\ensuremath{\PLambda}}\xspace}

\def\LorLbar     {\kern \thebaroffset\optbar{\kern -\thebaroffset \PLambda}\xspace}

\def\Xires       {{\ensuremath{\PXi}}\xspace}

\def\Omegares    {{\ensuremath{\POmega}}\xspace}

\def\Lc          {{\ensuremath{\Lz^+_\cquark}}\xspace}
\def\Lcp          {{\ensuremath{\Lz^+_\cquark}}\xspace}

\def\Xicp        {{\ensuremath{\Xires^+_\cquark}}\xspace}

\def\Xiccpp      {{\ensuremath{\Xires^{++}_{\cquark\cquark}}}\xspace}

\def\Lb           {{\ensuremath{\Lz^0_\bquark}}\xspace}

\def\BF         {{\ensuremath{\mathcal{B}}}\xspace}
\def\BR         {\BF}

\def\to                 {\ensuremath{\rightarrow}\xspace}

\def\AT#1     {\ensuremath{A_{\mathrm{T}}^{#1}}\xspace}           % 2

\def\C#1      {\ensuremath{\mathcal{C}_{#1}}\xspace}                       % 9
\def\Cp#1     {\ensuremath{\mathcal{C}_{#1}^{'}}\xspace}                    % 7
\def\Ceff#1   {\ensuremath{\mathcal{C}_{#1}^{\mathrm{(eff)}}}\xspace}        % 9  
\def\Cpeff#1  {\ensuremath{\mathcal{C}_{#1}^{'\mathrm{(eff)}}}\xspace}       % 7
\def\Ope#1    {\ensuremath{\mathcal{O}_{#1}}\xspace}                       % 2
\def\Opep#1   {\ensuremath{\mathcal{O}_{#1}^{'}}\xspace}                    % 7

\newcommand{\nospaceunit}[1]{\ensuremath{\text{#1}}}       
\newcommand{\aunit}[1]{\ensuremath{\text{\,#1}}}       
\newcommand{\tev}{\aunit{Te\kern -0.1em V}\xspace}
\newcommand{\gev}{\aunit{Ge\kern -0.1em V}\xspace}
\newcommand{\mev}{\aunit{Me\kern -0.1em V}\xspace}
\newcommand{\kev}{\aunit{ke\kern -0.1em V}\xspace}
\newcommand{\ev}{\aunit{e\kern -0.1em V}\xspace}
 
\newcommand{\mevc}{\ensuremath{\aunit{Me\kern -0.1em V\!/}c}\xspace}
\newcommand{\gevc}{\ensuremath{\aunit{Ge\kern -0.1em V\!/}c}\xspace}
\newcommand{\mevcc}{\ensuremath{\aunit{Me\kern -0.1em V\!/}c^2}\xspace}
\newcommand{\gevcc}{\ensuremath{\aunit{Ge\kern -0.1em V\!/}c^2}\xspace}
\def\mum  {\ensuremath{\,\mu\nospaceunit{m}}\xspace}

\def\nb {\aunit{nb}\xspace}

\def\fb   {\ensuremath{\aunit{fb}}\xspace}
\def\invfb   {\ensuremath{\fb^{-1}}\xspace}

\def\ps   {\ensuremath{\aunit{ps}}\xspace}

\newcommand{\chisq}{\ensuremath{\chi^2}\xspace}

\newcommand{\chisqip}{\ensuremath{\chi^2_{\text{IP}}}\xspace}

\def\gsim{{~\raise.15em\hbox{$>$}\kern-.85em
          \lower.35em\hbox{$\sim$}~}\xspace}
\def\lsim{{~\raise.15em\hbox{$<$}\kern-.85em
          \lower.35em\hbox{$\sim$}~}\xspace}

\def\sPlot{\mbox{\em sPlot}\xspace}

\def\sqs   {\ensuremath{\protect\sqrt{s}}\xspace}

\def\pt         {\ensuremath{p_{\mathrm{T}}}\xspace}

\def\ptot       {\ensuremath{p}\xspace}
\def\et         {\ensuremath{E_{\mathrm{T}}}\xspace}

\def\mrad{\aunit{mrad}\xspace}

\def\evtgen     {\mbox{\textsc{EvtGen}}\xspace}

\def\geant      {\mbox{\textsc{Geant4}}\xspace}

\def\photos     {\mbox{\textsc{Photos}}\xspace}

\def\pythia     {\mbox{\textsc{Pythia}}\xspace}

\def\tell1  {TELL1\xspace}
\def\ukl1   {UKL1\xspace}

\usepackage{cite} % Allows for ranges in citations
\usepackage{mciteplus}
\def\Omegabcn{{\ensuremath{\Omegares_{\bquark\cquark}^{0}}}\xspace}
\def\Xibcn{{\ensuremath{\Xires_{\bquark\cquark}^{0}}}\xspace}
\def\Hbcn{{\ensuremath{H_{\bquark\cquark}^{0}}}\xspace}
\def\Xibn{{\ensuremath{\Xires_{\bquark}^{0}}}\xspace}
\def\Xibcp{{\ensuremath{\Xires_{\bquark\cquark}^{+}}}\xspace}
\def\Xiccpp{{\ensuremath{\Xires_{\cquark\cquark}^{++}}}\xspace}

\def\XibcnXicdecay{{\ensuremath{\Xibcn\to\Xicp\pim}}\xspace}
\def\XibnXicdecay{{\ensuremath{\Xibn\to\Xicp\pim}}\xspace}
\def\XibcnLcdecay{{\ensuremath{\Xibcn\to\Lcp\pim}}\xspace}
\def\LbnLcdecay{{\ensuremath{\Lb\to\Lcp\pim}}\xspace}
\def\OmegabcnXicdecay{{\ensuremath{\Omegabcn\to\Xicp\pim}}\xspace}
\def\OmegabcnLcdecay{{\ensuremath{\Omegabcn\to\Lcp\pim}}\xspace}

\def\HbcnXicdecay{{\ensuremath{\Hbcn \to\Xicp\pim}}\xspace}
\def\HbcnLcdecay{{\ensuremath{\Hbcn \to\Lcp\pim}}\xspace}

\def\XibnXicFullDecay{{\ensuremath{\Xibn\to\Xicp(\to\ p \Km \pip)\pim}}\xspace}

\def\LbnLcFullDecay{{\ensuremath{\Lb\to\Lc(\to\ p \Km \pip)\pim}}\xspace}

\def\HbcnXicFullDecay{{\ensuremath{\Hbcn\to\Xicp(\to\ p \Km \pip)\pim}}\xspace}
\def\HbcnLcFullDecay{{\ensuremath{\Hbcn\to\Lc(\to\ p \Km \pip)\pim}}\xspace}

\def\Xicp{{\ensuremath{\Xires_{\cquark}^{+}}}\xspace}

\def\cls{\ensuremath${\rm CL}_{s}$\xspace}

\newcommand{\xx}{\ensuremath{\kern 0.5em }}

\def\sPlot{\mbox{\em sPlot}\xspace}

\def\genxicctwo   {\mbox{\textsc{GenXicc2.0}}\xspace}

\def\roostats     {\mbox{\textsc{RooStats}}\xspace}

\usepackage{multirow} % for complicated table
\usepackage{booktabs} % for complicated table
\usepackage{rotating}

\usepackage{longtable} % only for template; not usually to be used in PAPERs

\begin{document}

%%%%%%%%%%%%%%%%%%%%%%%%%
%%%%% Title     %%%%%%%%%
%%%%%%%%%%%%%%%%%%%%%%%%%
\renewcommand{\thefootnote}{\fnsymbol{footnote}}
\setcounter{footnote}{1}

% %%%%%%% CHOOSE TITLE PAGE--------
%\onecolumn
%\input{title-LHCb-INT}
%\input{title-LHCb-ANA}
%\input{title-LHCb-CONF}
%\input{title-LHCb-FIGURE}
% ===============================================================================
% Purpose: LHCb-PAPER journal paper title page template
% Author: 
% Created on: 2010-09-25
% ===============================================================================

%%%%%%%%%%%%%%%%%%%%%%%%%
%%%%%  TITLE PAGE  %%%%%%
%%%%%%%%%%%%%%%%%%%%%%%%%
\begin{titlepage}
\pagenumbering{roman}

% Header ---------------------------------------------------
\vspace*{-1.5cm}
\centerline{\large EUROPEAN ORGANIZATION FOR NUCLEAR RESEARCH (CERN)}
\vspace*{1.5cm}
\noindent
\begin{tabular*}{\linewidth}{lc@{\extracolsep{\fill}}r@{\extracolsep{0pt}}}
\ifthenelse{\boolean{pdflatex}}% Logo format choice
{\vspace*{-1.5cm}\mbox{\!\!\!\includegraphics[width=.14\textwidth]{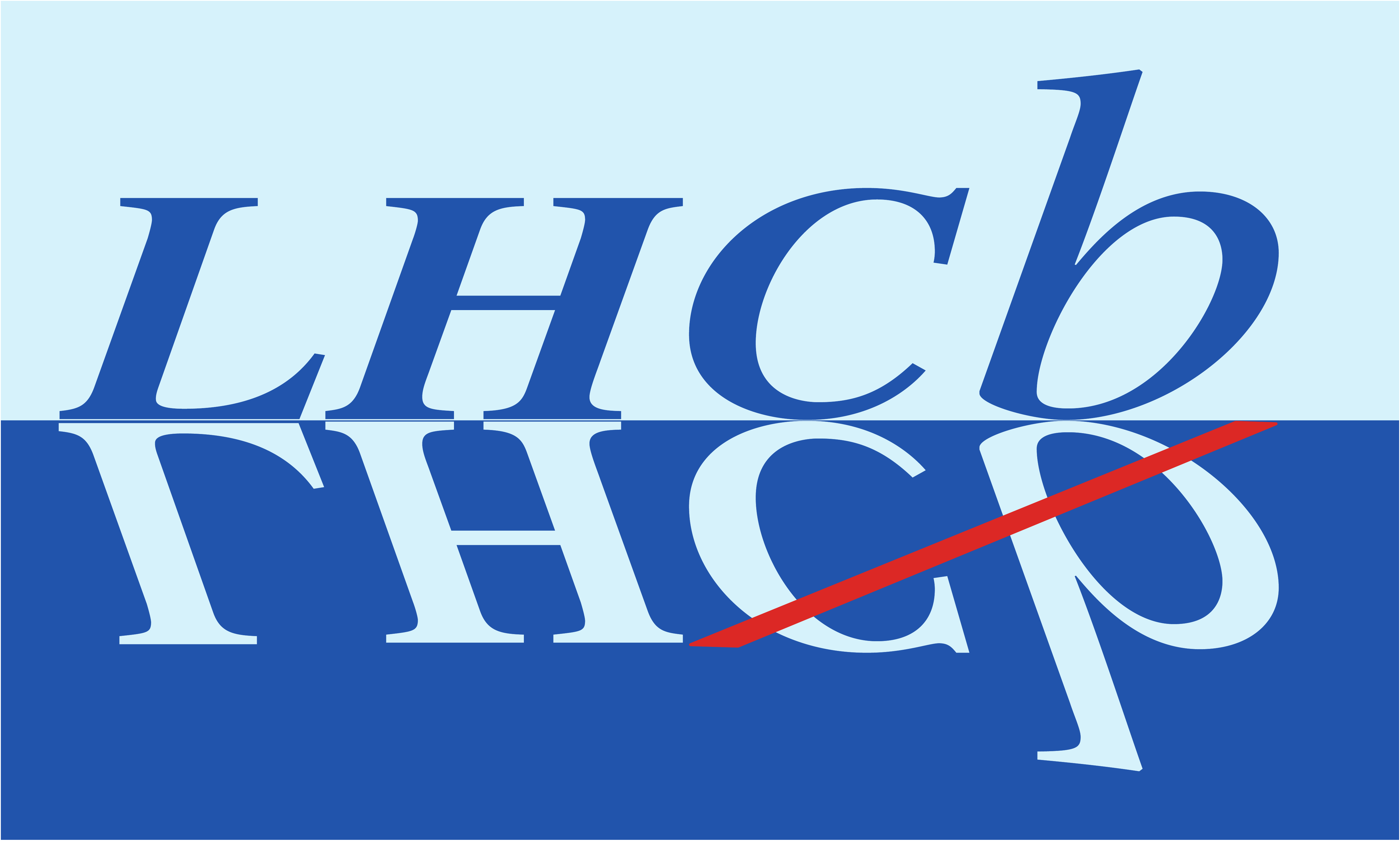}} & &}%
{\vspace*{-1.2cm}\mbox{\!\!\!\includegraphics[width=.12\textwidth]{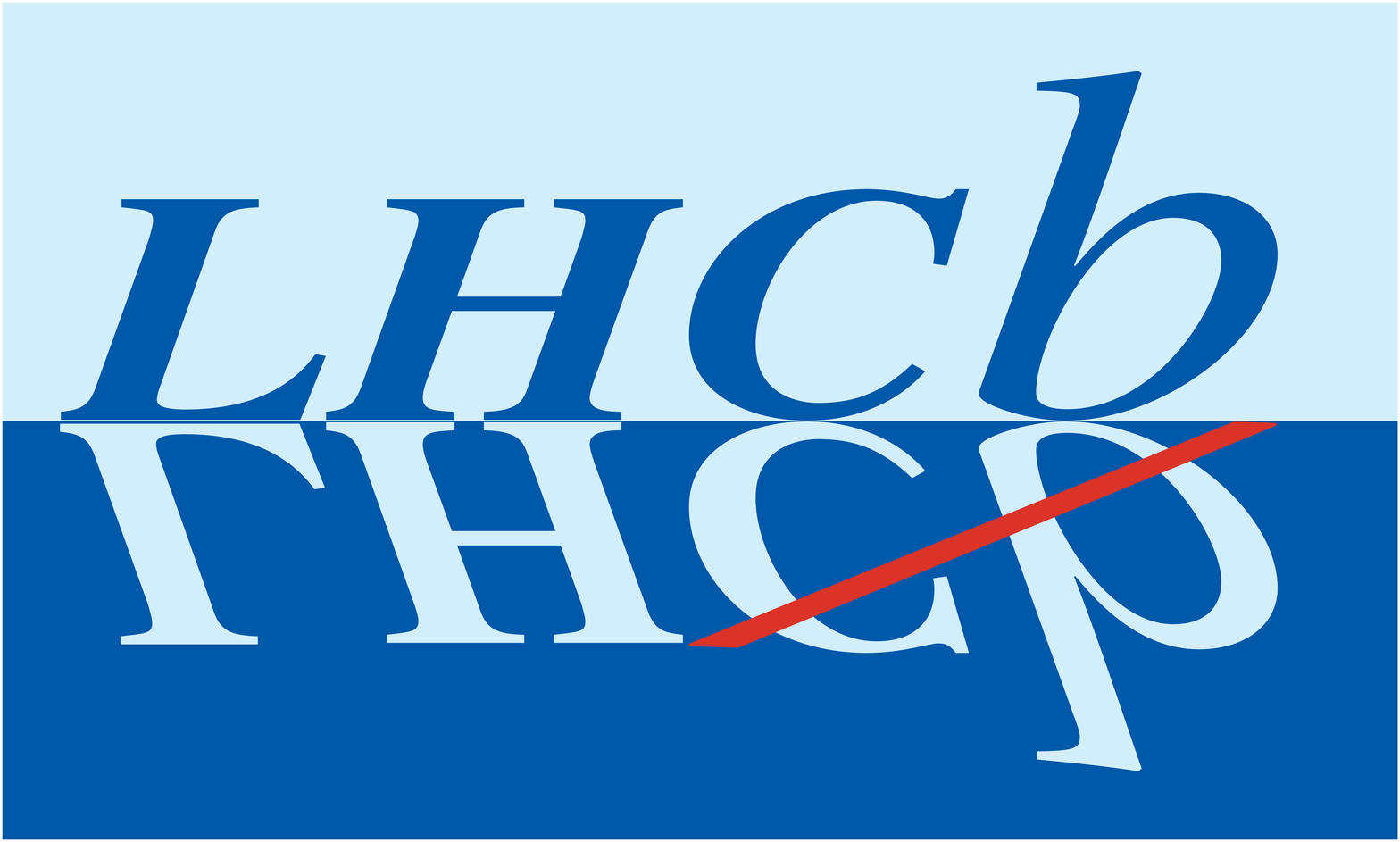}} & &}%
\\
 & & CERN-EP-2021-046 \\  % ID 
 & & LHCb-PAPER-2021-002 \\  % ID 
 & & Sep 13, 2021\\
%& & \today \\ % Date - Can also hardwire e.g.: 23 March 2010
% & & v0.2 \\
 & & \\
% not in paper \hline
\end{tabular*}

\vspace*{4.0cm}

% Title --------------------------------------------------
{\normalfont\bfseries\boldmath\huge
\begin{center}
% DO NOT EDIT HERE. Instead edit macro in main.tex to keep metadata correct
  \papertitle 
\end{center}
}

\vspace*{1.0cm}

% Authors -------------------------------------------------
\begin{center}
%In the footnote, replace 'paper' by 'Letter' in case of submission to PRL or PLB 
% Edit macro in main.tex to keep metadata correct
\paperauthors\footnote{Authors are listed at the end of this paper.}
\end{center}

\vspace{\fill}

% Abstract -----------------------------------------------
\begin{abstract}
  \noindent
 The first search for the doubly heavy \Omegabcn baryon and a search for the
 \Xibcn baryon are performed using $pp$ collision data collected via the \lhcb experiment from
  2016 to 2018 at a centre-of-mass energy of $13\tev$, corresponding to an integrated luminosity of 5.2\invfb. 
  The baryons are reconstructed via their decays to $\Lc \pim$ and $\Xicp \pim$.
  No significant excess is found for invariant masses between 6700 and 7300\mevcc, in a rapidity range from 2.0 to 4.5 and a transverse momentum range from 2 to 20\gevc.
  Upper limits are set on the ratio of the \Omegabcn~and~\Xibcn production cross-section times the branching fraction to $\Lc\pim$~(\Xicp\pim) relative to that of the $\Lb$~($\Xibn$) baryon, for different lifetime hypotheses,
  at $95\%$ confidence level.
  The upper limits range from $0.5\times10^{-4}$ to $2.5\times10^{-4}$ for the \mbox{\OmegabcnLcdecay} (\mbox{\XibcnLcdecay}) decay,
  and from $1.4\times10^{-3}$ to $6.9\times10^{-3}$ for the \mbox{\OmegabcnXicdecay} (\mbox{\XibcnXicdecay}) decay, depending on the considered mass and lifetime of the \Omegabcn (\Xibcn) baryon. 
\end{abstract}

\vspace*{1.5cm}

\begin{center}
  Published in Chin. Phys. C45 (2021) 093002
%JHEP 
%  Nature~Physics /
%  sciPost~Physics /
%  J. Instr. /
%  Instruments 
\end{center}

\vspace{\fill}

{\footnotesize 
% Edit macro in main.tex to keep metadata correct
\centerline{\copyright~\papercopyright. \href{\paperlicenceurl}{\paperlicence}.}}
\vspace*{2mm}

\end{titlepage}

%%%%%%%%%%%%%%%%%%%%%%%%%%%%%%%%
%%%%%  EOD OF TITLE PAGE  %%%%%%
%%%%%%%%%%%%%%%%%%%%%%%%%%%%%%%%

%  empty page follows the title page ----
\newpage
\setcounter{page}{2}
\mbox{~}
%\newpage
%
%% Author List ----------------------------
%%  You need to get a new author list!
%\input{LHCb_authorlist.tex}
%
%The author list for journal publications is provided by the Membership Committee shortly after 'approval to go to paper' has been given.
%%It will be made available on the page
%%\verb!http://www.physik.uzh.ch/~strauman/forMemCo/LHCb-PAPER-XXXX-XXX/! .
%It will be sent to you by email shortly after a paper number has beens assigned.
%The author list should be included already at first circulation, 
%to allow new members of the collaboration to verify whether they have been included correctly.
%Occasionally a misspelled name is corrected or associated institutions become full members.
%In that case, a new author list will be sent to you.
%In case line numbering doesn't work well after including the authorlist, try moving the \verb!\bigskip! after the last author to a separate line.
%
%
%The authorship for Conference Reports should be ``The LHCb
%  collaboration'', with a footnote giving the name(s) of the contact
%  author(s), but without the full list of collaboration names.

%\twocolumn
% %%%%%%%%%%%%% ---------

\renewcommand{\thefootnote}{\arabic{footnote}}
\setcounter{footnote}{0}

%%%%%%%%%%%%%%%%%%%%%%%%%%%%%%%%
%%%%%  Table of Content   %%%%%%
%%%%%%%%%%%%%%%%%%%%%%%%%%%%%%%%
%%%% Uncomment if desired
%\tableofcontents
\cleardoublepage

%%%%%%%%%%%%%%%%%%%%%%%%%
%%%%% Main text %%%%%%%%%
%%%%%%%%%%%%%%%%%%%%%%%%%

\pagestyle{plain} % restore page numbers for the main text
\setcounter{page}{1}
\pagenumbering{arabic}

%% Uncomment during review phase. 
%% Comment before a final submission.
%\linenumbers

%% This is the main body
%% It is useful to have a single file so comemnts are not missed in overleaf.
\section{Introduction}
\label{sec:Introduction}
The constituent quark model was initially proposed by 
Murray Gell-Mann~\cite{GellMann:1964nj} and George Zweig~\cite{Zweig:1964jf} 
for classification of hadrons formed from light quarks (\uquark, \dquark, \squark) 
and understanding their quantum numbers. It was later extended to hadrons containing heavy \cquark or \bquark quarks~\cite{Fleck:1988vm}. 
In addition to baryons containing a single heavy quark, 
the theory also predicts baryons comprising two heavy quarks. 
Such doubly heavy baryons provide a unique probe for quantum chromodynamics, 
the gauge theory of strong interactions.
In 2017, the \lhcb collaboration reported the first observation of the \Xiccpp
baryon containing two charm quarks through the decay
\mbox{$\Xiccpp\to \Lcp\Km\pip\pip$}~\cite{LHCb-PAPER-2017-018}.\footnote{The inclusion of charge-conjugate modes is implied throughout this paper.} 
The $\Xiccpp$ state was later confirmed in the decay to $\Xicp\pip$~\cite{LHCb-PAPER-2018-026}. Its lifetime, mass and production 
cross-section were subsequently measured~\cite{LHCb-PAPER-2018-019,LHCb-PAPER-2019-035,LHCb-PAPER-2019-037}. 
To date, no baryons containing one \bquark and one \cquark quark, or two \bquark quarks, have been observed experimentally. 
An observation would enrich our knowledge of baryon spectroscopy and improve our understanding of the quark structure inside baryons.

The ground-state baryons containing one \bquark and one \cquark quark, the \Omegabcn~(\bquark\cquark\squark), \Xibcn~(\bquark\cquark\dquark) 
and \Xibcp~(\bquark\cquark\uquark) states, 
have been considered within various theoretical models.
Most studies predict the masses of the \Omegabcn and \Xibcn baryons 
to be between 6700 and 7200\mevcc~\cite{Bagan:1992za,Roncaglia:1995az,
Lichtenberg:1995kg,Ebert:1996ec,
SilvestreBrac:1996wp,Kiselev:2001fw,
Anikeev:2001rk,
Ebert:2002ig,He:2004px,Albertus:2006ya,
Roberts:2007ni,
Zhang:2008rt,Gerasyuta:2008zy,Azizi:2014jxa,
Karliner:2014gca,Shah:2017liu,Brown:2014ena}.

The lifetime of the \Omegabcn baryon is predicted to be $0.22\pm0.04\ps$~\cite{Kiselev:2001fw}, while the lifetime of \Xibcn is predicted to be in the range of 0.09 to 
0.28\ps~\cite{Kiselev:2001fw,Karliner:2014gca,Kiselev:1999kh,Cheng:2019sxr}.
The production cross-section of the \Xibcn baryon in proton-proton ($\proton\proton$) 
collisions at a centre-of-mass energy $\sqs=14\tev$ is expected to 
lie in the range between 19 to 39\nb, derived from Ref.~\cite{Zhang:2011hi}, in the \Xibcn pseudorapidity ($\eta$)
range of $1.9 <\eta< 4.9$, depending on the required minimum value of the momentum component transverse to the beam direction~(\pt)
of the \Xibcn particle.
Recently, the \lhcb experiment found no significant \Xibcn baryon signal in the predicted
mass range
using the \mbox{$\Xibcn\to D^0 p\Km$}
decay mode~\cite{LHCb-PAPER-2020-014}.

\begin{figure}[!htbp]
\begin{center}
\includegraphics[width=0.45\textwidth]{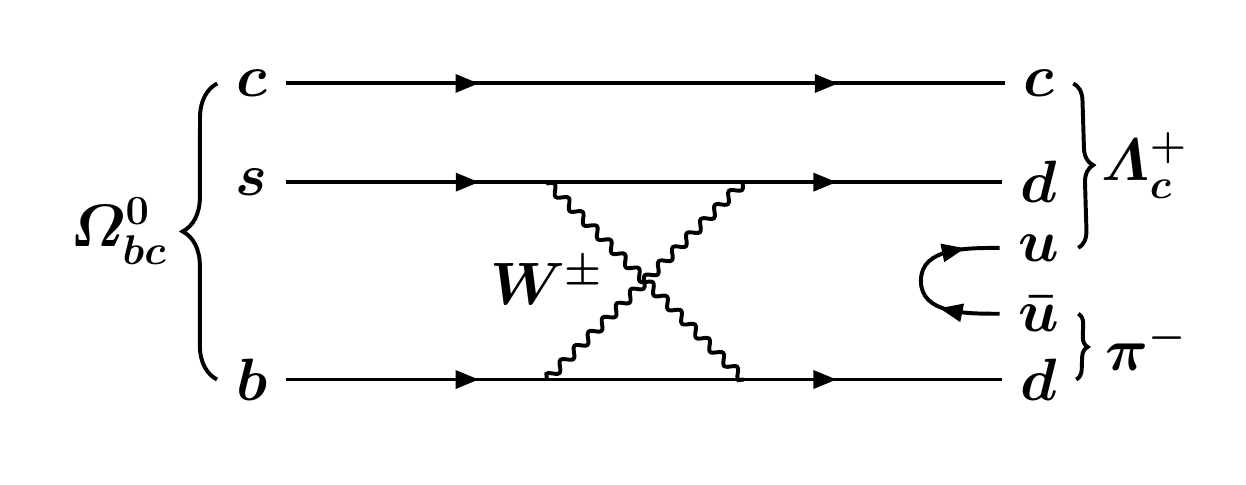}
\includegraphics[width=0.45\textwidth]{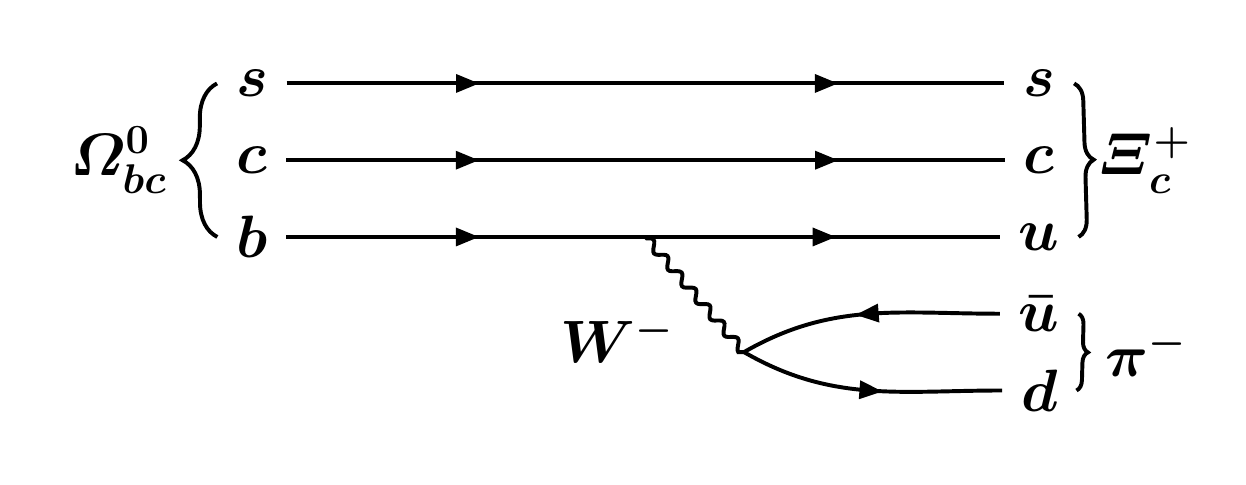}
\includegraphics[width=0.45\textwidth]{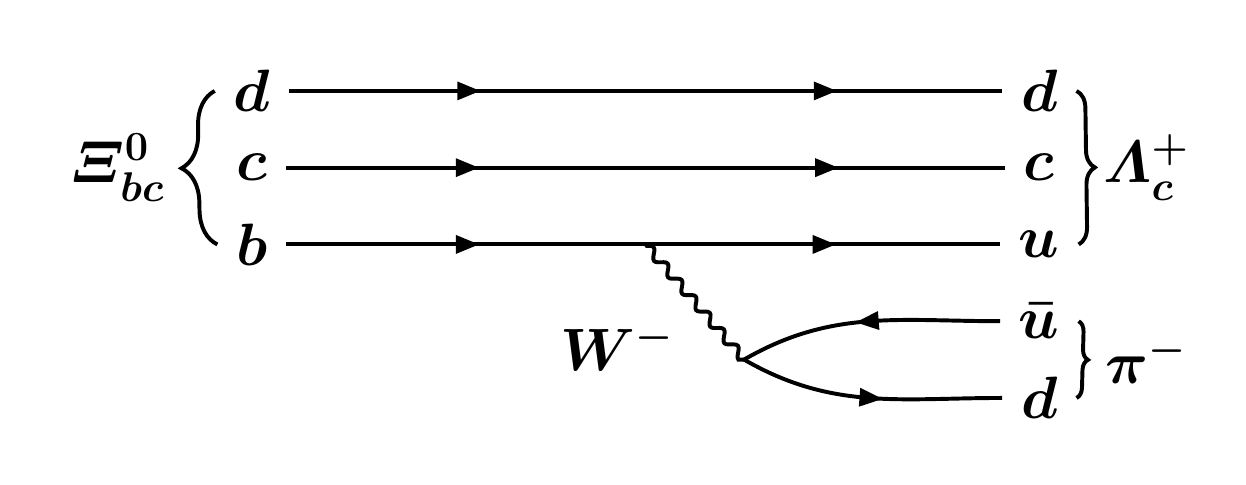}
\includegraphics[width=0.45\textwidth]{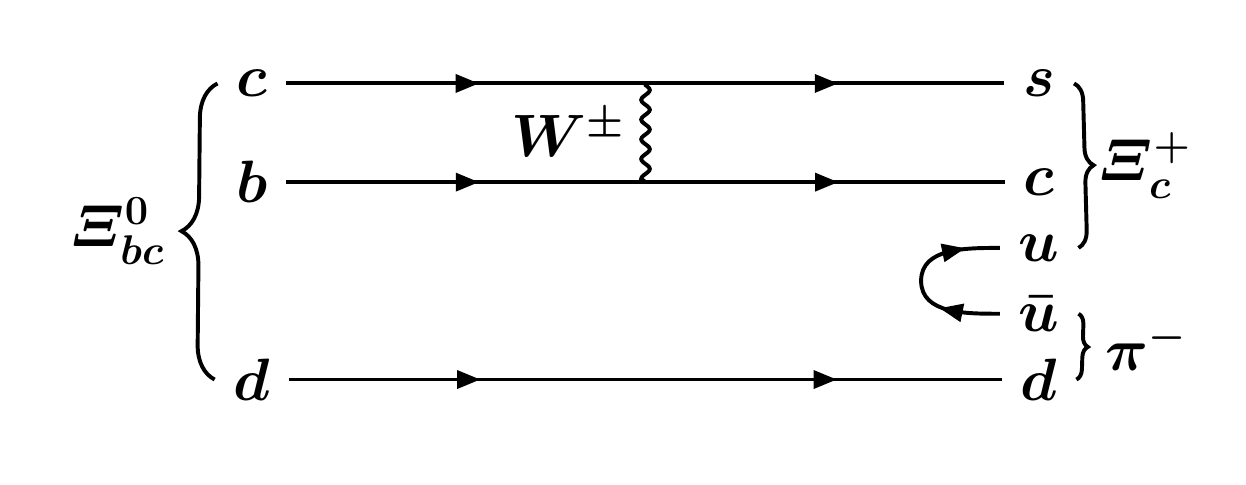}
\end{center}
\caption{Examples of Feynman diagrams for the $\OmegabcnLcdecay$, $\OmegabcnXicdecay$, $\XibcnLcdecay$ and $\XibcnXicdecay$ decays.}
\label{fig:FeynmanDiagram}
\end{figure}

This article reports the first search for the \Omegabcn baryon and 
a new search for the \Xibcn baryon, both via decay chains $\Lc\pim$ with 
\mbox{$\Lc\to\proton\Km\pip$} or $\Xicp\pim$ with 
\mbox{$\Xicp\to\proton\Km\pip$} in the \lhcb experiment.
Examples of Feynman diagram of the four signal decay modes are shown in Fig.~\ref{fig:FeynmanDiagram}.
There are few theoretical predictions on the branching fractions of these decay modes. Ref.~\cite{Wang:2017mqp} predicts the branching fractions of the \mbox{\OmegabcnXicdecay} and \mbox{\XibcnLcdecay} decays to be $1.6\times10^{-7}$ and $3.0\times10^{-7}$, 
respectively. However, uncertainties are not quoted.
Inputs from experimental studies are necessary to deepen our understanding of the properties of \Omegabcn and \Xibcn baryons, and can provide valuable reference for future searches.
Besides, the distinct experimental signatures of these decays make it promising to search for them in the \lhcb experiment considering the high detection efficiency of the \lhcb detector.

The \Omegabcn and \Xibcn baryons are not differentiated and are collectively 
denoted as \Hbcn hereafter, unless otherwise stated.
The production cross-section times branching fraction of \mbox{\HbcnLcdecay~(\HbcnXicdecay)} decay is measured relative to that of
the control channel \mbox{\LbnLcdecay (\XibnXicdecay)}.
This takes advantage of identical final-state particles and similar topology.
The search is performed in the mass range between $6700$ and $7300\mevcc$ using 
the $\proton\proton$ collision data collected with the \lhcb experiment 
at $\sqs=13\tev$,
corresponding to an integrated luminosity of $5.2\invfb$. The \Hbcn baryons are reconstructed in the fiducial region of rapidity~($y$) between $2.0$ 
and $4.5$ and with \pt between $2$ and $20\gevc$.

%%%%%% %%%%%% %%%%%% %%%%%% %%%%%% %%%%%% %%%%%% %%%%%% %%%%%% %%%%%% %%%%%% %%%%%% %%%%%% %%%%%% %%%%%%
%%%%%%%%%%%%%%%%%%%%%%%%%%%%%%%%%%%%%%%%%%%%%%%%%%%%%%%%%%%%%%%%%%%%%%%%%%%%%%%%%%%%%%%%%%%%%%%%%
\section{Detector and simulation}

The \lhcb detector~\cite{LHCb-DP-2008-001,LHCb-DP-2014-002} is a single-arm forward
spectrometer covering the \mbox{pseudorapidity} range $2<\eta <5$,
designed for the study of particles containing \bquark or \cquark
quarks. The detector includes a high-precision tracking system
consisting of a silicon-strip vertex detector surrounding the $pp$
interaction region~\cite{LHCb-DP-2014-001}, a large-area silicon-strip detector located
upstream of a dipole magnet with a bending power of approximately
$4{\mathrm{\,Tm}}$, and three stations of silicon-strip detectors and straw
drift tubes~\cite{LHCb-DP-2017-001} placed downstream of the magnet.
The tracking system provides a measurement of the momentum, \ptot, of charged particles with
a relative uncertainty that varies from 0.5\% at low momentum to 1.0\% at 200\gevc.
The minimum distance of a track to a primary $pp$ collision vertex (PV), the impact parameter (IP), 
is measured with a resolution of $(15+29/\pt)\mum$,
where \pt is in\,\gevc.
Different types of charged hadrons are distinguished using information
from two ring-imaging Cherenkov detectors~\cite{LHCb-DP-2012-003}. 
Photons, electrons, and hadrons are identified by a calorimeter system consisting of
scintillating-pad and preshower detectors, and an electromagnetic
% calorimeter
and a hadronic calorimeter. 
Muons are identified by a
system composed of alternating layers of iron and multiwire
proportional chambers~\cite{LHCb-DP-2012-002}.
The online event selection is performed by a trigger~\cite{LHCb-DP-2012-004}, 
which consists of a hardware stage, based on information from the calorimeter and muon
systems, followed by a software stage, which applies a full event
reconstruction.
At the hardware trigger stage, events are required to have at least one hadron with \et larger than 3.5\gev.
The software trigger requires a two-, three- or four-track
secondary vertex with a significant displacement from any PV. At least one charged particle
must have $\pt > 1.6\gevc$ and be
inconsistent with originating from any PV.
A multivariate algorithm~\cite{BBDT,LHCb-PROC-2015-018} is used for
the identification of secondary vertices consistent with the decay
of a \bquark hadron.

Simulated samples are produced to model the effects of the detector acceptance and the
imposed selection requirements.
In the simulation, $pp$ collisions are generated using
\pythia~\cite{Sjostrand:2007gs, Sjostrand:2006za} 
with a specific \lhcb configuration~\cite{LHCb-PROC-2010-056}.
A dedicated generator, \genxicctwo, 
is used to simulate the \Hbcn baryon production~\cite{Chang:2009va},
with the mass and lifetime of the \Hbcn baryon
set to $m(\Hbcn)=6900\mevcc$ and $\tau(\Hbcn)=0.4\ps$.
Simulation samples with different mass (6700--7300\mevcc) and lifetime (0.2--0.4\ps) hypotheses are obtained using a weighting technique with the generator level information on signal.
Decays of unstable particles
are described by \evtgen~\cite{Lange:2001uf}, in which final-state
radiation is generated using \photos~\cite{Davidson:2010ew}.
The interaction of the generated particles with the detector, and its response,
is implemented using the \geant toolkit~\cite{Allison:2006ve, Agostinelli:2002hh} 
as described in Ref.~\cite{LHCb-PROC-2011-006}. 
For the two control channels, \mbox{\LbnLcdecay} and \mbox{\XibnXicdecay}, 
\pythia is used to simulate the $pp$ collisions and the production of the \Lb and \Xibn baryons.

%%%%%%%%%%%%%%%%%%%%%%%%%%%%%%%%%%%%%%%%%%%%%%%%%%%%%%%%%%%%%%%%%%%%%%%%%%%%%%%%%%%%%%%%%%%%%%%%%
\section{Reconstruction and selection}
For both the \Hbcn signal and the control channels,
 the \mbox{$\Lcp, \Xicp\to p\Km \pip$} candidates are reconstructed from three charged particles identified as a proton, kaon and pion, respectively.
The tracks are required to have good quality,
and to be inconsistent with 
originating from any PV in the event.
The tracks must also form a common vertex of good fit quality.
The \Lcp (\Xicp) candidate is required to have an invariant mass in the range 2271--2301\mevcc 
(2450--2488\mevcc), corresponding to approximately six times the \Lcp (\Xicp) mass resolution, 
and to be inconsistent with originating from any PV.
In the sample of selected \Lcp (\Xicp) candidates, there is a sizable background contamination 
from decays of other particles, such as $\Dp$ ($\Dsp$) decays to $\Km\pip\pip$ ($\Km\Kp\pip$) with a charged pion (kaon) misidentified as a proton, 
and background from $\phi\pip$ combinations where in \mbox{$\phi\to\Km\Kp$} decays a kaon is misidentified as a proton.
Such background candidates are rejected if the $\Km\pip\pip$, $\Km\Kp\pip$ or $\Km\Kp$ invariant mass is consistent with the known $\Dp$, $\Dsp$ or $\phi$ mass~\cite{PDG2020}, respectively,
when a charged pion (kaon) hypothesis is assigned to the proton candidate.

An additional charged particle identified as a pion and, with \pt greater than 0.2\gevc, is combined with the \Lcp (\Xicp) candidate to form an \Hbcn candidate. 
The \Hbcn candidates must have a vertex with good fit quality, a decay time larger than 0.05\ps, a \pt greater than 2\gevc
and a scalar sum of the \pt of the final-state particles greater than 5\gevc. 
Furthermore, the \Hbcn candidates are required to
be consistent with originating from a PV. To avoid contributions from duplicate tracks, the selected candidates are rejected if the angle
between any pair of the final-state particle tracks with same charge is smaller than 0.5\mrad. 

A boosted decision tree~(BDT) classifier~\cite{Breiman,AdaBoost} 
implemented in the TMVA toolkit~\cite{Hocker:2007ht,TMVA4} is used to further suppress combinatorial background.
A simulated signal sample in the mass range 6846--6954\mevcc and a background sample formed by candidates in an upper mass sideband region (7500--9000\mevcc) are used to train the BDT classifier.
Four different categories of variable are used as the BDT input. 
The first category exploits the non-zero lifetime of \Hbcn baryons and a displacement of their vertices from any PV in the event. The variables comprise the 
\chisqip of all final-state particles forming the \Lcp (\Xicp) and \Hbcn candidates with respect to their associated PV,
where \chisqip is defined as the difference in the vertex-fit \chisq of a given PV reconstructed with and without the particle under consideration, and the associated PV is the one with respect to which the \Hbcn candidate has the smallest \chisqip;
the sum of \chisqip of the four final-state particles; 
and $\chisq$ of the flight distance of the \Lcp (\Xicp) and \Hbcn candidates.
The second category consists of kinematic variables, including
\pt of the final-state particles and the \Lcp (\Xicp) and \Hbcn candidates, and
the angle between the \Hbcn momentum vector and the displacement vector pointing from the associated PV to the \Hbcn decay
vertex. The third category comprises the vertex-fit $\chisq$ of the \Lcp (\Xicp) and \Hbcn candidates,
and $\chisq$ of a kinematic fit~\cite{Hulsbergen:2005pu} of the signal decay chain constraining the \Hbcn candidate to originate from the associated PV. 
The fourth category consists of identification variables of the final-state particles.

The BDT threshold is chosen to maximize the figure of merit,
\mbox{$\varepsilon/(\tfrac{\alpha}{2} + \sqrt{B})$}~\cite{Punzi:2003bu}.
Here, $\varepsilon$ is the selection efficiency of signal candidates determined from simulation, 
$B$ is the expected background number in the signal mass region, and $\alpha=5$ is the signal significance.
This threshold is estimated using the signal sample simulated with the default mass and lifetime values. 
The same selection is applied to the control modes.

%%%%%%%%%%%%%%%%%%%%%%%%%%%%%%%%%%%%%%%%%%%%%%%%%%%%%%%%%%%%%%%%%%%%%%%%%%%%%%%%%%%%%%%%%%%%%%%%%
\section{Yield determination}
\begin{figure}[!htbp]
      \centering
      \includegraphics[width=0.45\textwidth]{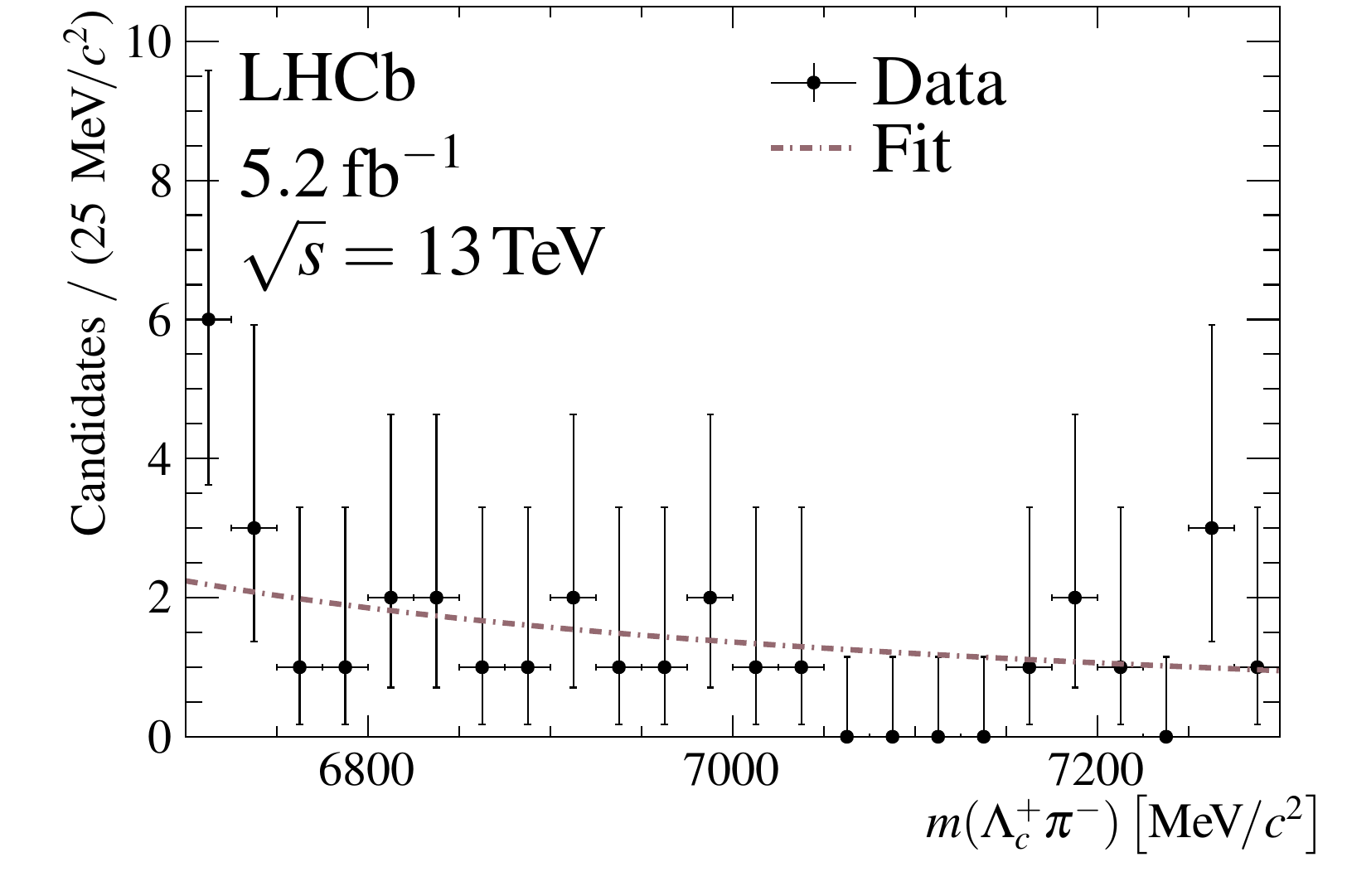}
      \includegraphics[width=0.45\textwidth]{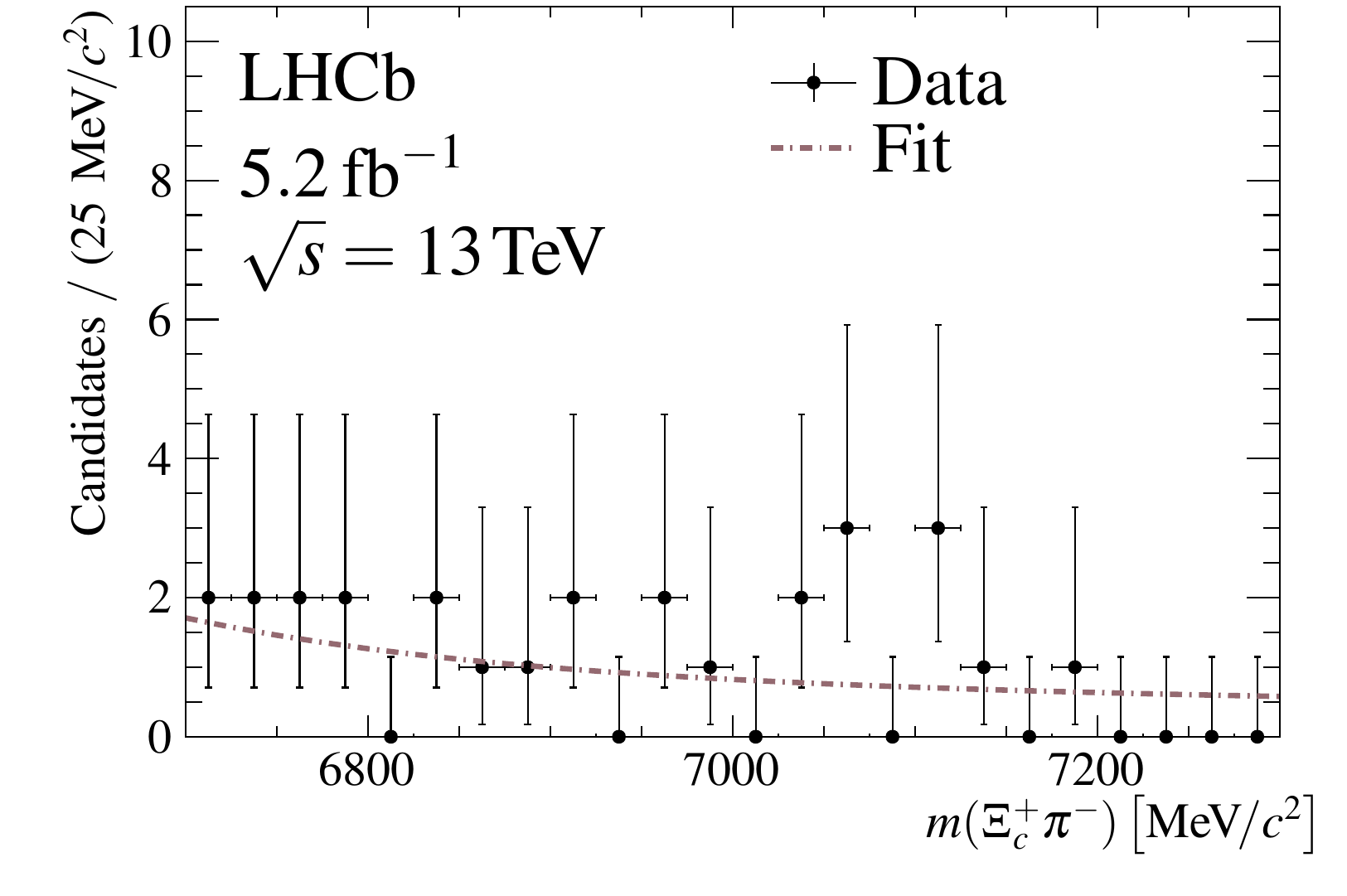}
      \caption{Invariant mass distributions of selected (left) \HbcnLcdecay and (right) \HbcnXicdecay candidates (black points), together with results of the background only fit (brown dashed line).
      } 
      \label{fig:Threeyears_unblind_signal}
  \end{figure}

To improve the resolution of the mass of the \Hbcn candidates, 
the $\Lcp\pim$~($\Xicp\pim$) invariant mass is calculated after constraining the
$\Lcp$~(\Xicp) mass to its known value~\cite{PDG2020} and requiring the 
\Hbcn candidate to be consistent with originating from its associated PV.
The obtained invariant mass distributions of \Hbcn candidates, $m(\Lcp\pim)$ and $m(\Xicp\pim)$, are shown 
in Fig.~\ref{fig:Threeyears_unblind_signal}.
To search for the \Hbcn signals, the mass distributions are fitted using an unbinned maximum-likelihood fit.
A double-sided Crystal Ball function~\cite{Skwarnicki:1986xj} is used to model the signal,
with tail parameters fixed from simulation, and the peak position and width allowed to vary in the fit.
The background shape is interpolated from a double-exponential fit to a lower (6100--6650\mevcc) and an upper (7500--9000\mevcc) sideband region of the $\Lcp\pim$~($\Xicp\pim$) mass distribution.
No significant excess is observed across the searched mass range.

\begin{figure}[!htbp]
\centering
\includegraphics[width=0.45\textwidth]{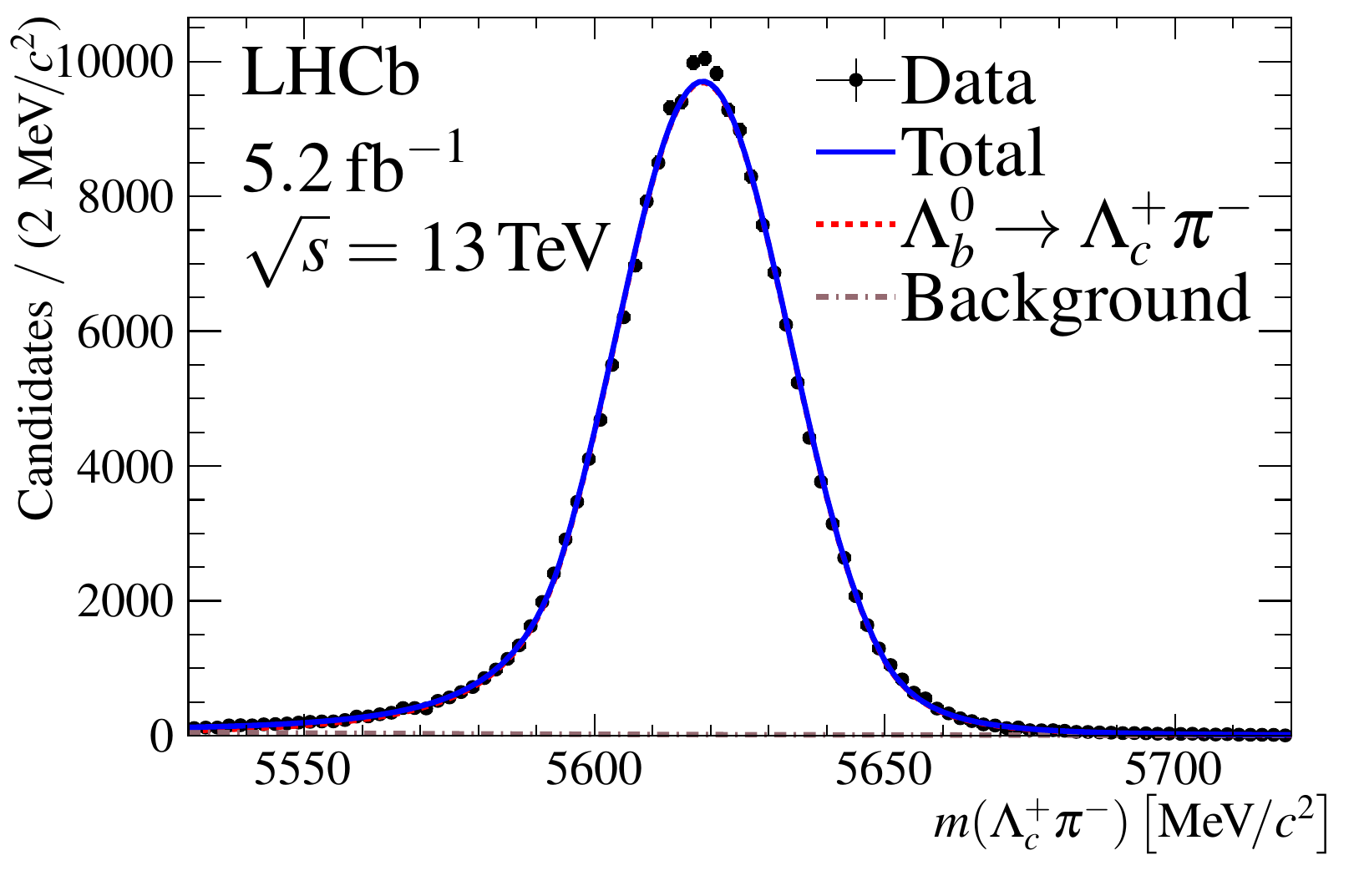}
\includegraphics[width=0.45\textwidth]{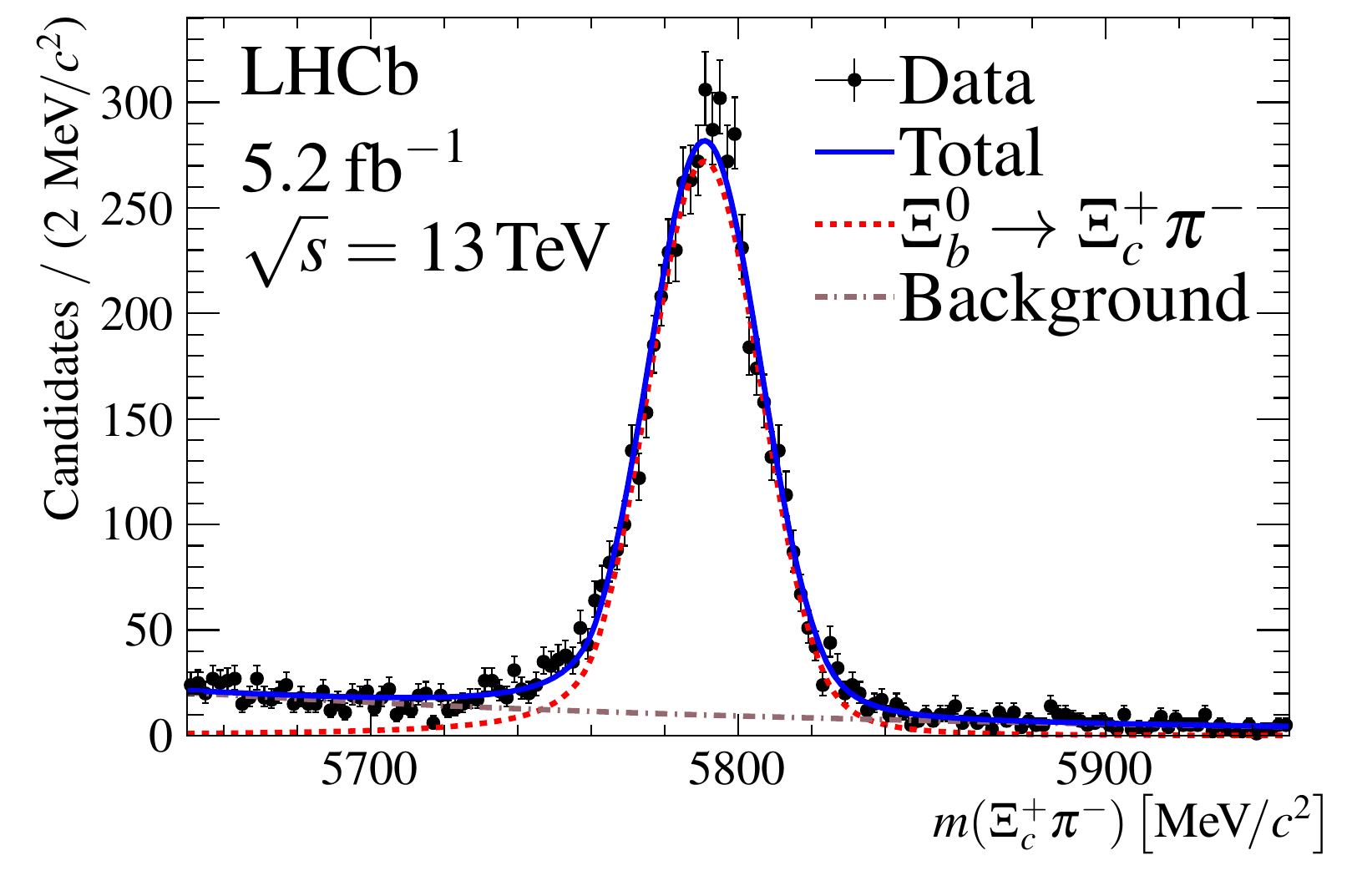}
\caption{Invariant mass distributions of (left) \mbox{\LbnLcFullDecay} and (right) \mbox{\XibnXicFullDecay} candidates with the fit results overlaid (blue solid line). The black points represent the data, the red dashed line represents the signal contribution, and the gray dashed line represents the combinatorial background. }
\label{fig:lckpipimassRunII}
\end{figure}

The $\Lcp\pim$ and $\Xicp\pim$ invariant mass distributions of the selected \mbox{\LbnLcFullDecay} and \mbox{\XibnXicFullDecay} candidates
are shown in Fig.~\ref{fig:lckpipimassRunII}.
The yields are obtained from unbinned maximum-likelihood fits to the invariant mass distributions,
using the fit model described above.
The yields are determined to be $191\,000\pm 500$ and $5\,490\pm 80$ for \mbox{\LbnLcFullDecay} and \mbox{\XibnXicFullDecay}, respectively. 

%%%%%%%%%%%%%%%%%%%%%%%%%%%%%%%%%%%%%%%%%%%%%%%%%%%%%%%%%%%%%%%%%%%%%%%%%%%%%%%%%%%%%%%%%%%%%%%

%%%%%%%%%%%%%%%%%%%%%%%%%%%%%%%%%%%%%%%%%%%%%%%%%%%%%%%%%%%%%%%%%%%%%%%%%%%%%%%%%%%%%%%%%%%%%%%%%
\section{Ratio of production cross-sections}

The ratio $\cal{R}$ of the \Hbcn baryon production cross-section multiplied by the branching fraction of the \mbox{$\HbcnLcdecay$ (\HbcnXicdecay)} decay relative to that of the $\Lb$ ($\Xibn$) baryon can be written as 
\begin{eqnarray}
\label{eq:br}
  {\cal{R}}(\Lcp\pim)&\equiv&\frac{\sigma(pp \to \Hbcn X)\ \BR\left( \HbcnLcFullDecay \right)}{\sigma(pp \to \Lb X)\ \BR\left( \LbnLcFullDecay \right)}\\ \nonumber
  &=& \frac{N(\HbcnLcdecay)}{N(\LbnLcdecay)}\cdot \frac{\varepsilon(\Lb)}{\varepsilon(\Hbcn)}, \\ 
  {\cal{R}}(\Xicp\pim)&\equiv&\frac{\sigma(pp \to \Hbcn X)\ \BR\left( \HbcnXicFullDecay \right)}{\sigma(pp \to \Xibn X)\ \BR\left( \XibnXicFullDecay \right)}\\ \nonumber
  &=& \frac{N(\HbcnXicdecay)}{N(\XibnXicdecay)}\cdot \frac{\varepsilon(\Xibn)}{\varepsilon(\Hbcn)},
\end{eqnarray}
where $N$ and $\varepsilon$ are the signal yield and the efficiency for 
the corresponding decay modes, respectively The efficiency accounts for the 
geometrical acceptance, trigger, reconstruction, and event selection.
The $\cal{R}$ is determined in the fiducial region $2<y<4.5$ and $2<\pt<20\gevc$.

Efficiencies are determined from the simulated samples. 
The \pt distributions of \Lb and \Xibn baryons are not well modeled in simulation. 
To improve the description, a gradient boosted weighting method~\cite{Rogozhnikov:2016bdp} is used to apply a kinematic correction on the \pt distributions of the \Lb and \Xibn decay products of the simulated control samples. With this correction, a good agreement on the \Lb and \Xibn \pt distribution is seen between the data and simulation.
The track detection and 
particle identification efficiencies are calibrated with the data~\cite{LHCB-DP-2013-002,LHCB-PUB-2016-021,LHCb-DP-2018-001}.
The imperfect modeling of input variables used in the BDT training can bias the efficiency estimation. 
To suppress such effects, 
ratios between the BDT response distribution of the background-subtracted data sample and that of the simulated sample are calculated using the control channel.
The background subtraction is performed using the \sPlot method~\cite{Pivk:2004ty} with $m(\Lcp\pim)$ and $m(\Xicp\pim)$ as discriminating variables.
These ratios are applied as correction weights to the simulated samples for all reconstructed decay modes. 

The total efficiency ratio $\varepsilon(\Lb)/\varepsilon(\Hbcn)$ is determined to be $3.18\pm0.05$, 
and $\varepsilon(\Xibn)/\varepsilon(\Hbcn)$ is calculated to be $3.00\pm0.02$ for
$m(\Hbcn)=6900\mevcc$ and $\tau(\Hbcn)=0.4\ps$.
The efficiency is larger for the control mode, mainly due to the longer lifetime of the \Lb and \Xibn baryons.
The efficiency depends on the mass and lifetime hypotheses of the \Hbcn state, and is evaluated from simulation.
The kinematic properties of the fully simulated samples are weighted to match those of the generator-level sample to calculate the efficiency for different \Hbcn mass and lifetime assumptions.

%%%%%%%%%%%%%%%%%%%%%%%%%%%%%%%%%%%%%%%%%%%%%%%%%%%%%%%%%%%%%%%%%%%%%%%%%%%%%%%%%%%%%%%%%%%%%%%%%
\section{Systematic uncertainties}
Various sources of systematic uncertainties on ${\cal{R}}$ are estimated and combined in quadrature. 
The effect of imperfect description of the mass distributions on the yield estimates is studied using alternative signal and background models. For the signal model, the Hypatia~\cite{Santos:2013gra} function is used instead of the nominal double-sided Crystal Ball function.
For the background model of the control modes, the nominal double-exponential function is 
replaced by a first-order polynomial function.
As the background model for the \Hbcn decay modes is interpolated from the sidebands,
its uncertainty is evaluated by both replacing the nominal function with an exponential function and varying the sideband regions.
The largest deviation with respect to the nominal result is taken as the corresponding uncertainty.
In total, the associated systematic uncertainty is estimated to be $0.1\%$ and $0.9\%$ for \mbox{\HbcnLcdecay} and \mbox{\HbcnXicdecay} decays, respectively.

In the ${\cal{R}}$ ratios, 
systematic uncertainties arising from the track detection efficiency largely cancel,
and the uncertainty due to limited size of simulation samples is determined to be
1.6\% (0.7\%) on ${\cal{R}} (\Lcp\pim)$ (${\cal{R}} (\Xicp\pim)$).
The particle identification efficiency is determined in bins of particle momentum, pseudorapidity,
and track multiplicity using control channels in the data~\cite{LHCb-DP-2018-001}. 
As the particle identification variables have large
dependencies on the momentum of the final-state particles, there are sizeable
differences in these efficiencies between the control and signal channels,
which do not cancel in the ratio measurement. 
Systematic effects arising from the choice of binning scheme are evaluated by varying the bin sizes and reevaluating the efficiency. 
The largest deviations from the nominal result, 1.7\% and 2.1\%, 
are assigned as the systematic uncertainty for the \mbox{\HbcnLcdecay} and \mbox{\HbcnXicdecay} decays, respectively.

The \Lcp (\Xicp) mass resolution shows a difference between data and simulation, which affects the selection efficiency. 
It results in a 0.2\% systematic uncertainty contribution for the
\mbox{\HbcnXicdecay} decay,
while the contribution for the \mbox{\HbcnLcdecay} decay is below 0.1\%.
This systematic uncertainty is negligible compared to other sources.

The imperfect simulation of the signal and control modes are considered by applying corrections to the BDT response and kinematic properties of the simulated control mode samples. To assess the systematic uncertainty in these corrections, the correcting weights are varied within their uncertainties. 
The largest deviation from the nominal result is taken as the systematic uncertainty.
Combining the uncertainties from the BDT response correction and the kinematic modeling of
the simulated control samples gives an uncertainty of 1.6\% for the \mbox{\HbcnLcdecay} channel, and 3.0\% for the \mbox{\HbcnXicdecay} channel. 
The analysis relies on the \Xibcn \pt model implemented in simulation. No systematic uncertainty is assigned to this model. 

The algorithm used to compute the \chisqip was updated during data collection,
which causes a mismatch between data and simulation and introduces systematic effects in the efficiency estimation.
The corresponding uncertainty was found to be $5\%$ in the previous $\Xibcn$ search~\cite{LHCb-PAPER-2020-014}.
Checks by varying the \chisqip-related requirements show that the uncertainty well covers the change of result.
Therefore, a $5\%$ systematic uncertainty is assigned.

The systematic uncertainties are summarized 
in Table~\ref{tab:sys_err}. 
The total systematic uncertainty is $5.7\%$ for \mbox{\HbcnLcdecay} 
and $6.3\%$ for \mbox{\HbcnXicdecay}, for a \Hbcn mass of 6900\mevcc and lifetime of 0.4\ps. 
These values of systematic uncertainties are also used for other assumed lifetime and mass hypotheses.

\begin{table}[!htbp]
    \caption{Sources of systematic uncertainty obtained for an \Hbcn mass of 6900\mevcc and lifetime of 0.4\ps. The total is the quadratic sum of the individual systematic uncertainties.}
\centering
\begin{tabular}{lcc}
\hline
 & \HbcnLcdecay &\HbcnXicdecay \\ \hline
Fit model & $0.1\%$ & $0.9\%$ \\
Size of simulated samples & 1.6\% & 0.7\% \\
Particle identification efficiency  & $1.7\%$ & $2.1\%$ \\
Mass resolution  &  $<0.1\%~~~$            & $0.2\%$              \\
Simulation model  & $1.6\%$ & $3.0\%$ \\
\chisqip simulation & 5.0\% & 5.0\% \\ \hline
Total &   $5.7\%$           &  $6.3\%$   \\
\hline
\end{tabular}\label{tab:sys_err}
\end{table}

%%%%%%%%%%%%%%%%%%%%%%%%%%%%%%%%%%%%%%%%%%%%%%%%%%%%%%%%%%%%%%%%%%%%%%%%%%%%%%%%%%%%%%%%%%%%%%%%%
\section{Results}
\begin{figure}[!htbp]
      \centering
      \includegraphics[width=0.45\textwidth]{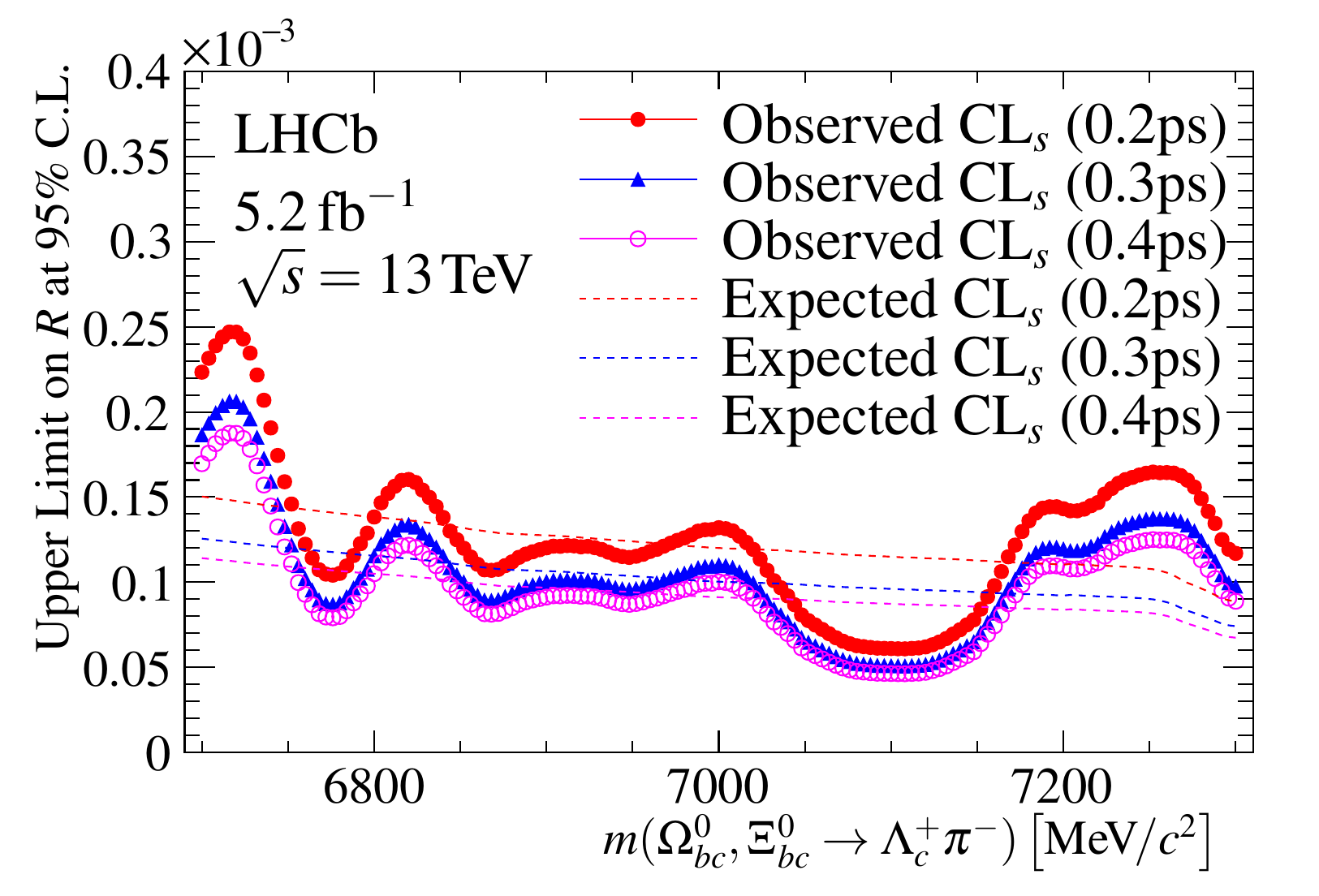}
      \includegraphics[width=0.45\textwidth]{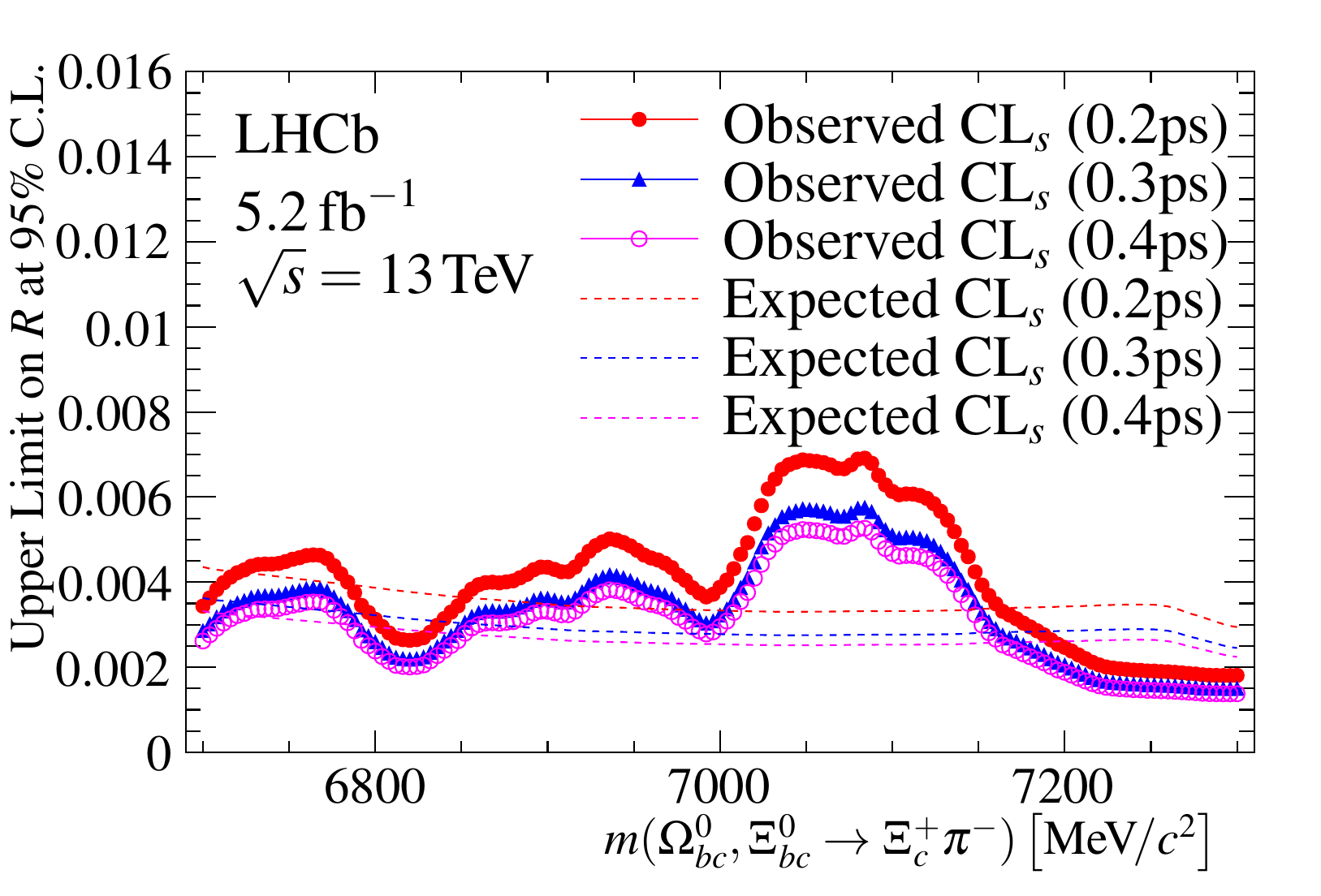}
      \caption{Upper limits (dotted lines) on the ratio of production cross-section for \Omegabcn and \Xibcn via decays to (left) \Lcp\pim and (right) \Xicp\pim to that of control channels \mbox{\LbnLcdecay} and \mbox{\XibnXicdecay}. The dotted (dashed) colored lines represent the observed (expected) upper limits. The assumed lifetime hypotheses for the \Omegabcn (\Xibcn) are 0.2\ps (red filled circles), 0.3\ps (blue triangles), and 0.4\ps (magenta open circles).}
      \label{fig:Threeyears_upperlimit_lifetime}
  \end{figure}

No evidence for a \Omegabcn or a \Xibcn baryon is observed in the inspected mass range. Upper limits are set 
at 95\% confidence level on the ratios ${\cal{R}}(\Lcp\pim)$ and ${\cal{R}}(\Xicp\pim)$ under different mass and lifetime hypotheses for the $\Omegabcn$ and $\Xibcn$ baryons,
using the asymptotic \cls method implemented in the \roostats framework~\cite{Moneta:2010pm,Cowan:2010js} considering the systematic uncertainties.
The assumed masses of the $\Omegabcn$ and $\Xibcn$ baryons are varied from 6700 to 7300\mevcc with a 
step size of 4\mevcc, and the lifetime values of 0.2\ps, 0.3\ps, and 0.4\ps are considered.
The calculated upper limits are shown in Fig.~\ref{fig:Threeyears_upperlimit_lifetime}, as a function of the \Hbcn mass. These results are obtained for the sum of the \Omegabcn and \Xibcn production and as such hold for the two individually.

%%%%%%%%%%%%%%%%%%%%%%%%%%%%%%%%%%%%%%%%%%%%%%%%%%%%%%%%%%%%%%%%%%%%%%%%%%%%%%%%%%%%%%%%%%%%%%%%%
\section{Conclusions}

The first search for the doubly heavy baryon \Omegabcn and a new search for the \Xibcn baryon in the mass range from 6700 to 7300\mevcc are presented, using $pp$ collision data
collected at a centre-of-mass energy $\sqs=13\tev$ with the \lhcb experiment. 
The data set corresponds to an integrated luminosity of 5.2\invfb.
The \Omegabcn (\Xibcn) baryon is reconstructed in the $\Lcp\pim$ and $\Xicp\pim$ decay modes. No evidence of a signal %(below two standard deviation) 
is found. 
Upper limits at 95\% confidence level on the ratio of the \Omegabcn~(\Xibcn) production cross-section times its branching fraction relative to that of the $\Lb$ (\Xibn) baryon
are calculated in the rapidity range $2.0 < y < 4.5$ and transverse momentum range $2<\pt < 20\gevc$
under different \Omegabcn~(\Xibcn) mass and lifetime hypotheses.
The upper limits range from $0.5\times10^{-4}$ to $2.5\times10^{-4}$ for the \mbox{\OmegabcnLcdecay (\XibcnLcdecay)} decay,
and from $1.4\times10^{-3}$ to $6.9\times10^{-3}$ for the \mbox{\OmegabcnXicdecay (\XibcnXicdecay)} decay, for the considered lifetime and mass hypotheses. These results constitute the first limit on the production of the \Omegabcn baryon. Further measurements will be possible with the larger data samples expected at the upgraded \lhcb experiments~\cite{Bediaga:2018lhg} and with additional decay modes.

%%%%%%%%%%%%%%%%%%%%%%%%%%%%%%%%%%%%%%%%%%%%%%%%%%%%%%%%%%%%%%%%%%%%%%%%%%%%%%%%%%%%%%%%%%%%%%%%%
%%%%%%%%%%%%%%%%%%%%%%%%%%%%%%%%%%%%%%%%
%%%%%%%%%%%%%%%%%%%%%%%%%%%%%%%%%%%%%%%%
%%%%%%%%%%%%%%%%%%%%%%%%%%%%%%%%%%%%%%%%%%%%%%%%%%%%%%%%%%%%%%%%%%%%%%%%%%%%%%%%%%%%%%%%%%%%%%%%

% Do not include this in any draft (just for information in the template)
% \section{Acknowledgements paragraph}

% Include the following text in the Acknowledgements section in all paper
% drafts. It is not needed for analysis notes or conference reports.

% The text below are the acknowledgements as approved by the collaboration
% board. Extending the acknowledgements to include individuals from outside the
% collaboration who have contributed to the analysis should be approved by the
% EB. The extra acknowledgements are normally placed before the standard 
% acknowledgements, unless it matches better with the text of the standard 
% acknowledgements to put them elsewhere. They should be included in the draft 
% for the first circulation. Except in exceptional circumstances, to be approved by the
% EB chair, authors of the paper should not be named in extended acknowledgements.

% %\vspace{1cm}

\section*{Acknowledgements}
%
% These Acknowledgements valid from 3-May-2019
%
\noindent We express our gratitude to our colleagues in the CERN
accelerator departments for the excellent performance of the LHC. We
thank the technical and administrative staff at the LHCb
institutes.
We acknowledge support from CERN and from the national agencies:
CAPES, CNPq, FAPERJ and FINEP (Brazil); 
MOST and NSFC (China); 
CNRS/IN2P3 (France); 
BMBF, DFG and MPG (Germany); 
INFN (Italy); 
NWO (Netherlands); 
MNiSW and NCN (Poland); 
MEN/IFA (Romania); 
MSHE (Russia); 
MICINN (Spain); 
SNSF and SER (Switzerland); 
NASU (Ukraine); 
STFC (United Kingdom); 
DOE NP and NSF (USA).
We acknowledge the computing resources that are provided by CERN, IN2P3
(France), KIT and DESY (Germany), INFN (Italy), SURF (Netherlands),
PIC (Spain), GridPP (United Kingdom), RRCKI and Yandex
LLC (Russia), CSCS (Switzerland), IFIN-HH (Romania), CBPF (Brazil),
PL-GRID (Poland) and NERSC (USA).
We are indebted to the communities behind the multiple open-source
software packages on which we depend.
Individual groups or members have received support from
ARC and ARDC (Australia);
AvH Foundation (Germany);
EPLANET, Marie Sk\l{}odowska-Curie Actions and ERC (European Union);
A*MIDEX, ANR, Labex P2IO and OCEVU, and R\'{e}gion Auvergne-Rh\^{o}ne-Alpes (France);
Key Research Program of Frontier Sciences of CAS, CAS PIFI, CAS CCEPP, 
Fundamental Research Funds for the Central Universities, 
and Sci. \& Tech. Program of Guangzhou (China);
%Key Research Program of Frontier Sciences of CAS, CAS PIFI,
%Thousand Talents Program, and Sci. \& Tech. Program of Guangzhou (China);
RFBR, RSF and Yandex LLC (Russia);
GVA, XuntaGal and GENCAT (Spain);
the Leverhulme Trust, the Royal Society
 and UKRI (United Kingdom).

% Comment this in for paper drafts; do not include this in analysis note, conference and figure reports
%\input{acknowledgements}

%\input{supplementary}

%\input{appendix}

% This should be taken out in the final paper
%\input{supplementary-app}

\addcontentsline{toc}{section}{References}
%\setboolean{inbibliography}{true}
\bibliographystyle{LHCb}
\bibliography{main,standard,LHCb-PAPER,LHCb-CONF,LHCb-DP,LHCb-TDR,Xibc,myref}

\newpage
% LHCb collaboration author list
% Data extracted on April 1st, 2021 at 7:57pm for paper reference LHCb-PAPER-2021-002
\centerline
{\large\bf LHCb collaboration}
\begin
{flushleft}
\small
R.~Aaij$^{32}$,
C.~Abell{\'a}n~Beteta$^{50}$,
T.~Ackernley$^{60}$,
B.~Adeva$^{46}$,
M.~Adinolfi$^{54}$,
H.~Afsharnia$^{9}$,
C.A.~Aidala$^{85}$,
S.~Aiola$^{25}$,
Z.~Ajaltouni$^{9}$,
S.~Akar$^{65}$
% ...

{\footnotesize \it

% ...
$^{9}$Universit{\'e} Clermont Auvergne, CNRS/IN2P3, LPC, Clermont-Ferrand, France\\
% ...
$^{25}$INFN Sezione di Milano, Milano, Italy\\
% ...
$^{32}$Nikhef National Institute for Subatomic Physics, Amsterdam, Netherlands\\
% ...
$^{46}$Instituto Galego de F{\'\i}sica de Altas Enerx{\'\i}as (IGFAE), Universidade de Santiago de Compostela, Santiago de Compostela, Spain\\
% ...
$^{50}$Physik-Institut, Universit{\"a}t Z{\"u}rich, Z{\"u}rich, Switzerland\\
% ...
$^{54}$H.H. Wills Physics Laboratory, University of Bristol, Bristol, United Kingdom\\
% ...
$^{60}$Oliver Lodge Laboratory, University of Liverpool, Liverpool, United Kingdom\\
% ...
$^{65}$University of Cincinnati, Cincinnati, OH, United States\\
% ...
$^{68}$Syracuse University, Syracuse, NY, United States\\
% ...
$^{85}$University of Michigan, Ann Arbor, United States, associated to $^{68}$\\
\bigskip
% ...
\medskip
}
\end{flushleft}

% The author list for journal publications is generated from the
% Membership Database shortly after 'approval to go to paper' has been
% given.  It will be sent to you by email shortly after a paper number
% has been assigned.  The author list should be included in the draft used for 
% first and second circulation, to allow new members of the collaboration to verify
% that they have been included correctly. Occasionally a misspelled
% name is corrected, or associated institutions become full members.
% Therefore an updated author list will be sent to you after the final
% EB review of the paper.  In case line numbering doesn't work well
% after including the authorlist, try moving the \verb!\bigskip! after
% the last author to a separate line.

% The authorship for Conference Reports should be ``The LHCb
% collaboration'', with a footnote giving the name(s) of the contact
% author(s), but without the full list of collaboration names.

% The authorship for Figure Reports should be ``The LHCb
% collaboration'', with no contact author and without the full list 
% of collaboration names.

\end{document}